\documentclass[final,authoryear,5p]{elsarticle}

\usepackage{epsfig}

\usepackage{amssymb}
\usepackage{multirow}
\usepackage{amsmath}

\usepackage{pdflscape}

\usepackage[ps2pdf,%
a4paper=true,%
breaklinks=true,%
colorlinks=true,%
pdfauthor={First Author et al.},%
pdftitle={Template for manuscripts in Advances in Space Research}%
]{hyperref}

\journal{Advances in Space Research}

\begin{document}

\begin{frontmatter}



\title{Electron production by solar Ly-$\alpha$ line radiation in the ionospheric D-region }


\author[label1]{Aleksandra Nina\corref{cor}}
\address[label1]{Institute of Physics, University of Belgrade, Pregrevica 118,  Belgrade, Serbia}
\cortext[cor]{Corresponding author}
\ead{sandrast@ipb.ac.rs}


\author[label2]{Vladimir M. \v{C}ade\v{z}}
\ead{vcadez@aob.rs}

\address[label2]{Astronomical Observatory, Volgina 7, 11060 Belgrade, Serbia}

\begin{abstract}
The hydrogen Ly-$\alpha$ line has a dominant influence in photo-ionization processes
in the unperturbed terrestrial ionospheric D region. In this paper, we present a
procedure of calculating the rate of photo-ionization induced by Ly-$\alpha$ photons
based on relaxation of electron density after intensive perturbations like those caused by solar X flares.
This theory is applied to the ends of relaxation periods following three cases of solar X flares
from May 5, 2010, February 18, 2011 and March 24, 2011.
The necessary data on low
ionospheric plasma parameters were collected by the very low frequency (VLF)
radio-wave technics. The electron concentration is calculated from the amplitude
and phase of the VLF signal emitted by the DHO transmitter in Germany and
recorded by a receiver located in Serbia.
\end{abstract}


\end{frontmatter}

\parindent=0.5 cm

\section{Introduction}

As a part of terrestrial atmosphere, the ionosphere is under permanent variable
influences coming from outer space and Earth's atmosphere and lithosphere.
The non-periodic and sudden events, such as solar flares \citep{mcr04,nin12a,nin11,kol11},
coronal mass ejections \citep{bal08,boc03}, influences of the processes in remote parts of
the universe like supernova explosions followed by hard X and $\gamma$ radiation \citep{ina07},
lightnings \citep{vos98,col11}, and some processes in the Earth lithosphere like volcanic
eruptions and earthquakes \citep{uta11,hek05} temporarily induce
 space and time changeable ionospheric perturbations. The variability of affects caused
by permanent sources of perturbations, like the quiet Sun radiation
or cosmic rays, lies in variations of their initial intensity, of
attenuation during propagation to the considered location and, finally,
of the total cross-section for the analyzed
process. In addition to a purely scientific interest, studying the
effects of specific ionospheric plasma perturbers finds practical
applications, primarily in telecommunications. Namely, changes in
the signal characteristics caused by varying ionospheric plasma
conditions and compositions, require their predictions in
order to deal with disturbances in signal reception.

One of the most important external influences on the chemical
processes in the lowest ionospheric layer, called the D region, is coming
from the solar Ly-$\alpha$ line (121.6 nm) radiation  whose presence
is periodically intensified during the day. Generally, in local plasma
this line participates in several processes such as the oxygen and
water cluster dissociation, and chemistry of minor species such as water
vapor, ozone and nitric oxide \citep{woo00}. In this paper we are
interested in production of electrons in plasma located in the
middle part of the D region as a result of
NO molecule photo-ionization by Ly-$\alpha$ photons.
A very important significance of this process lies in the fact
that the formation of the D region in the daytime is
primarily a result of the photo-ionization by Ly-$\alpha$ photons
\citep{nic60}, and reversly, reduction of the incoming solar flux
including the Ly-$\alpha$ line is followed by disappearance
of the lowest ionospheric layer during the nighttime conditions.
The rate of this process depends on
the incident Ly-$\alpha$ flux, its attenuation during propagation
though higher atmospheric layers and on the NO density in local
plasma. As numerous studies have shown, all these parameters
are variable in space and time and they can be calculated from
experimental data obtained by various observational technics.

Data sets for the Ly-$\alpha$ irradiance are modeled based on
values recorded  by satellites such as, for example, the Atmospheric
Explorer E (AE-E), the Solar Mesospheric Explorer (SME), and the
Upper Atmosphere Research Satellite (UARS), and variations in the
Ly-$\alpha$ irradiance during solar cycles and seasons are presented
in \cite{woo00,fro09,cor11}.

The atmospheric Ly-$\alpha$ line absorbtion coefficients can be obtained
from data gathered by rockets, balloons, space shuttle and satellite
measurements, and the results exhibit a strong zenith
angle dependency which is analyzed in \cite{koc02}.
Also, measurements of the NO density
show values within a wide range at fixed altitudes
(\cite{aik64}; \cite{pea69}; \cite{bar12} and references therein).
As the incident flux of the Ly-$\alpha$ line, its absorption coefficient,
and the NO density at considered altitudes may have different values,
the resulting rate of relevant photo-ionization
process varies within 3 order of magnitudes \citep{aik69}.

In this work, we present a new procedure for determining the
photo-ionization rate induced by the Ly-$\alpha$ line in the middle parts
of the D region at altitudes between around 70 and 80 km
 during the relaxation period after larger perturbations such as solar flares. The basic
points of our approach are based on occurrence of transient ionospheric
perturbations resulting in electron density change and the fact
that the perturbed plasma
tends
to the original state
once the action of the perturber is over.

Generally, there are several global models that can be used to determine the electron density in the D region such as the International Reference
Ionosphere (IRI) model \citep{bil92} (an empirical model of unperturbed ionosphere that can be applied to the altitude range between 50 and 2000 km and that use rocket measurements as data base to determine the electron density in the D region), the Sodankyla Ion Chemistry (SIC) model \citep{tur92} (it was built to be a tool for interpretation of D-region incoherent scatter experiments and cosmic radio-noise absorption measurements and it can be run either as a steady-state model or a time-dependent model) and Mitra-Rowe chemical model \citep{mit72} in which electron density can be calculated by adjusting free parameters according to rocket-borne mass spectrometer measurements.

The suggested new theoretical procedure utilizes
data that can be obtained by
a continuous monitoring of the low ionosphere
using
primarily ground based technics: the radar and very low frequency (VLF) radio signal measurements.
The advantage of the latter is seen
in numerous transmitters and receivers located worldwide,
which provides a possibility to analyze a large part of the low ionosphere.
The theoretical equations are applied to
 examples of low ionospheric
perturbation taking three solar X flares as a convenient perturbing mechanism.
The electron density time and altitude dependencies are calculated  from
data obtained by the D region monitoring using VLF radio signals emitted by the
DHO transmitter located in Germany and recorded in Serbia by the AWESOME (Atmospheric Weather
Electromagnetic System for Observation Modeling and Education) receiver \citep{sch08}
within the Stanford/AWESOME Collaboration for Global VLF Research activities.

\section{Theoretical backgrounds}

Unlike the existing procedures for determining the Ly-$\alpha$ photo-ionization rate in
the ionosphere from calculated data related to the Ly-$\alpha$ radiation entering the terrestrial atmosphere,
its attenuation during propagation in the gas, and the NO density at locations of interest, in this
work we present a method that requires knowledge of
the electron density time variations
in the studied local medium
in the aftermath of an intensive perturbation when variations in ionization rate are not large.
This excludes periods of the local sunrise
and sunset, and time intervals characterized by large variations in photons and/or particle radiation
in the domain of sufficiently large energies to ionize some of species in the low ionospheric plasma.

As in the case of many similar studies like \cite{mit74} and \cite{mce78},
we start from the general form of the electron density dynamic:

  \begin{equation}
    \label{eq:1}
    {
    \frac{dN(\vec{r},t)}{dt}= {\cal G}(\vec{r},t)-{\cal L}(\vec{r},t),
    }
  \end{equation}
 where $N(\vec{r},t)$, ${\cal G}(\vec{r},t)$ and ${\cal L}(\vec{r},t)$ are electron density, electron gain and electron loss rates,
  respectively at location $\vec{r}$ and at time $t$. Here, the influence of transport processes is neglected as they become important
  at altitudes above 120-150 km \citep{bla06} and they are generally not included in models of the D region plasma \citep{tur92,wil78}.
  The above Eq.~(\ref{eq:1})
  is very complicated in its general form due to
  complexity of terms ${\cal G}(\vec{r},t)$ and ${\cal L}(\vec{r},t)$.
  For this reason, they can first be
  expressed in a form composed of quantities whose values can be obtained from
  observations and from modeling, which allows for calculations of the photo-ionization rate
  induced by the Ly-$\alpha$ line.
  In the case of absence of intensive sudden influences coming from the Earth,
  the dominant D region plasma ionization is caused by perturbers from the outer
  space.

   The contribution of these photons and particles on the low
  ionospheric plasma processes depends on their fluxes entering the atmosphere,
  attenuations during the propagation through higher atmospheric layers, and
  on the total ionization cross section at considered location.
As to the photo-ionization, the required
  wavelengths for
  the D region main species N$_2$, O$_2$, NO, and O are less than
  79.6, 102.7, 134.0, and 91.1 nm, respectively, and
 102.7 nm $\leq \lambda \leq$ 111.8 nm for the metastable O$_2$($^1\Delta_g$).
The variability of the electron production in the D layer thus depends on
  the type and intensity of the incoming radiation.
  For example, the hard X and gamma radiation penetrates the atmosphere with comparatively lower
  attenuation but their photo-ionization cross sections for the involved
   plasma constituents are much smaller than for other photons with smaller
   energy \citep{ber98}. Contrary, the large part of the UV domain has
   a strong attenuation in the higher atmospheric layers above the D region \citep{rat72}.
   Consequently, typical moderate intensity variations of these two types of radiation have little influence
   on the electron production in the D region.
In the case without sudden perturbations, the published investigations (for example \cite{mce78};
  \cite{tho74} and references therein)
show that  Ly-$\alpha$,
  cosmic, and the radiation at wavelength domain between 102.7 and 111.8 nm
  (these photons can ionize the metastable molecule O$_2$($^1\Delta_g$)) can be considered the main
  ionization sources.
  However,
  these studies show that the ionization rate
  induced by cosmic rays decreases
  with altitude and its influence can be negligible above about 70 km.
  Finally, the influence of radiation between 102.7 and 111.8 nm is
  smaller then that of Ly-$\alpha$
  photons at altitudes $h$ between 70 and 80 km.
  Consequently, we can assume that the ${\cal G}(\vec{r})$ is approximately
  equal to the photo-ionization rate induced in the D region by the
  ${\cal G}_{Ly\alpha}(\vec{r})$ radiation at considered heights:

  \begin{equation}
    \label{eq:2}
    {
    {\cal G}(\vec{r})={\cal G}_{Ly\alpha}(\vec{r}), \quad 70\rm km \leq h \leq 80\rm km.
    }
  \end{equation}
  Here, we assumed that time variation of
  the ionization rate is not significant and we can ignore the variable $t$.

  There are many processes in the ionosphere that result into electron density
  reduction. The rates of these processes depend on the structure and state of
  plasma in the considered medium in time and space. The recombination processes
  including the electron-ion, ion-ion and three body recombination have very important
  role so that the electron loss rate can be expressed as \citep{mit74,zig07}:

  \begin{equation}
    \label{eq:4}
    {
    {\cal L}(\vec{r},t)=\xi_{_L}(\vec{r},t)N^{2}(\vec{r},t),
    }
  \end{equation}
where $\xi_{_L}(\vec{r},t)$ is the effective recombination coefficient.

Finally, the Eqs~(\ref{eq:1})
and (\ref{eq:4}) give:

  \begin{equation}
    \label{eq:5}
    {
    \frac{dN(\vec{r},t)}{dt}= {\cal G}(\vec{r},t)-\xi_{_L}(\vec{r},t)N^{2}(\vec{r},t).
    }
  \end{equation}

The quantities in Eq.~(\ref{eq:5})
can be determined by combining the observational data, numerical modeling, and theoretical procedure.
In this sense, the electron density and its
time derivative can be calculated indirectly from numerically processed data recorded by
different technics like the radar and VLF radio signal monitoring of the low ionosphere
while the electron gain rate ${\cal G}(\vec{r},t)$ and coefficient $\xi_{_L}(\vec{r},t)$
will be determined by the theoretical procedure that follows.

In Eq.~(\ref{eq:2}), the Ly-$\alpha$ photo-ionization rate ${\cal
G}_{Ly\alpha}(\vec{r})$  which is related to conditionally unperturbed ionosphere,
can be derived from Eq.~(\ref{eq:5}) applied to the late stage of perturbation relaxation
in the aftermath of the flare. Thus, it is not necessary to know values of parameters in
Eq.~(\ref{eq:5}) during the entire flare period and the analysis can be restricted to
the late stage of plasma relaxation to the initial state
provided the large fluctuation due to Ly-alpha emission, and possible other perturbers like particle and electromagnetic (continuum and spectral lines radiation) are absent. In this case, the
variations of ${\cal G}(\vec{r},t)$ and $\xi_{_L}(\vec{r},t)$ become weakly time dependent
and we assume them approximately constant
within some finite time interval $\Delta t$:

\[
{\cal G}(\vec{r},t-\Delta t)\approx {\cal G}(\vec{r},t)\equiv \overline{{\cal G}}(\vec{r},t),
\]
\[\xi_{_L}(\vec{r},t-\Delta t)\approx \xi_{_L}(\vec{r},t)
\equiv\overline\xi_{_L}(\vec{r},t).
\]

Eq.~(\ref{eq:5}) can now be applied to the interval end points
 $t_{\rm 1}=t-\Delta t$ and $t_{\rm 2}=t$ which yields the following set
 of two algebraic equations for the unknown quantities $\overline{{\cal G}}(\vec{r},t)$ and
 $\overline \xi_{_L}(\vec{r},t)$:

\begin{equation}
\label{eq:9}
{
\begin{array}{l}
\displaystyle\left.\frac{dN}{dt}\right\vert_{\vec{r},t-\Delta t}=\overline{{\cal G}}(\vec{r},t)
-\overline \xi_{_L}(\vec{r},t) N^2(\vec{r},t-\Delta t)\\[.2cm]
\displaystyle\left.\frac{dN}{dt}\right\vert_{\vec{r},t}=\overline{{\cal G}}(\vec{r},t)
-\overline\xi_{_L}(\vec{r},t) N^2(\vec{r},t)
\end{array}
}
\end{equation}
which finally yields the expression for $\overline{{\cal G}}(\vec{r},t)$:

 {\begin{equation}
    \label{eq:13}
    {
\begin{array}{ll}
\displaystyle {\cal G}(\vec{r},t)=
\frac{N^2(\vec{r},t)\left.\frac{dN}{dt}\right\vert_{\vec{r},t-\Delta t}-N^2(\vec{r},t-\Delta t)\left.\frac{dN}{dt}\right\vert_{\vec{r},t}}{N^2(\vec{r},t)-N^2(\vec{r},t-\Delta t)}
\end{array}.
    }
  \end{equation}

The above analysis thus allows estimates of ${\cal G}_{Ly\alpha}(\vec{r})$ through
expressions Eq.~(\ref{eq:2}) and Eq.~(\ref{eq:13}). Here we wish to point out that the length of the relaxation period has to be large enough for the described procedure to be performed. As, on the other hand, this length depends on the intensity of the initial ionospheric perturbation, it means that only sufficiently intensive ionospheric perturbations (like solar flares class C or stronger) can be  treated by our analysis.


\section{Electron density modeling}

As in the case of numerous papers \citep{mcr00,gru08,zig07,tho11,nin11,nin12a,nin12b}, our calculation of electron density is based on the observed VLF signal data and their comparison with data obtained by simulations of the VLF signal propagation using the Long-Wave Propagation Capability (LWPC) numerical code (developed by the Naval Ocean Systems Center (NOSC), San Diego, USA \citep{fer98}). This numerical code uses
 Wait's model of vertically stratified ionosphere \citep{wai64} characterized by two independent parameters, the signal reflection hight $H'(t)$ and sharpness $\beta(t)$ (related to the electron density gradient as shown in \cite{nin12a}) which yields the following analytical expression for the electron density $N(h,t)$:
  \begin{equation}
    \label{eq:14}
    {
    N(h,t) = 1.43\cdot10^{13}e^{-\beta(t)H'(t)}e^{(\beta(t)-0.15)h}.
    }
  \end{equation}
Here, it should be noted that the application of this method to calculate the electron density assumes the spatial dependence of the quantities in Eq. (\ref{eq:13}) be reduced only to altitude $h$ according to the approximation of the low ionosphere stratification. Consequently, the forthcoming analysis requires data for signals that propagate through a medium whose plasma characteristics do not vary, at least locally, in the horizontal plane. This has to be taken into account i.e. signals from VLF transmitters located not too far from the receiver should be chosen in the analysis.

In the Eq.~\ref{eq:14}, both parameters $H'(t)$ and $\beta(t)$ are time dependent even if sudden ionospheric disturbances are absent. This, namely, comes from long time-dependent phenomena in the D-region like periodic variations in atmospheric composition, diurnal and seasonal variations of ionizing radiation intensity due to changes of solar zenith angle, and variations of radiation related to the solar cycle phase.
The dependence of $H'(t)$ and $\beta(t)$ on geographical position due to different local solar zenith angles is calculated in literature
for different periods of the solar cycle by, for example, detections of VLF radio atmospherics through VLF broadband observations \citep{han11}, and by comparisons of narrowband VLF data with simulations of signal propagation utilizing numerical codes such as LWPC \citep{tho93,mcr00}.
Because of numerous influences affecting the D region plasma, the parameters $H'(t)$ and $\beta(t)$ are shown to fall within intervals
70-75 km and 0.2-0.5 km$^{-1}$ during daytime in the case of quiet ionosphere \citep{fer98}.

In addition to differences in physical and chemical conditions explained above, the obtained discrepancies in calculated values result from application of different models like IRI \citep{bil92} and FIRI \citep{fri01}.
In calculations that include solar zenith dependence of Wait's parameters, the discrepancies among obtained $H'(t)$ and $\beta(t)$ are more pronounced at higher zenith angles. On the other hand, for smaller solar zenith angles, the time and space variations of plasma parameters are not so large and we can assume  $H'(t)$, $\beta(t)$ $\approx$ const for signals propagating along paths for which the solar zenith angle does not change much, and when the observation time is sufficiently short. This assumption of quasi constancy of $\beta$ and $H'$ will be applied in treatment that follows.


\section{Observed data and experimental setup}

We study the time period of ionospheric
relaxation in the aftermath of perturbations induced by
three solar X flares occurred on May 5$^{\rm {th}}$, 2010 (Case I),
February  18$^{\rm {th}}$, 2011 (Case II)
and March 24$^{\rm {th}}$, 2011 (Case III)
whose impacts on Earth's atmosphere are registered by
satellites
GOES-14 (Fig.~\ref{Fig1::signali}, top panel, Case I) and GOES-15 (Fig.~\ref{Fig1::signali}, middle and bottom panels, Case II and III)

\begin{figure}
\begin{center}
\includegraphics*[width=9cm,angle=0]{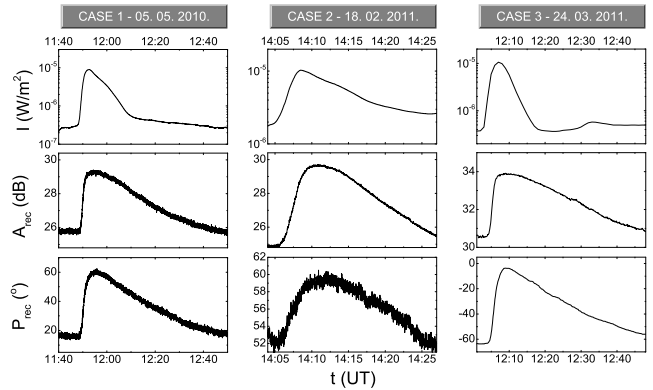}
\end{center}
\caption{Time dependencies of the intensity of X radiation registered by satellites
in the wavelength domain between 0.1 and 0.8 nm (top panels), and signal amplitude (middle panels) and phase (bottom panels) registered by AWESOME receiver located in Belgrade (Serbia) during presence three solar flares occurred on May 5$^{\rm {th}}$, 2010 (Case I), February  18$^{\rm {th}}$, 2011 (Case II), and March 24$^{\rm {th}}$, 2011 (Case III).}
\label{Fig1::signali}
\end{figure}

To calculate altitude and time dependencies of the electron density $N(h,t)$, we perform
the ground based low ionosphere monitoring using the 23.4 kHz VLF radio signal emitted by the DHO
transmitter in Rhauderfehn (Germany) and recorded by the AWESOME VLF receiver located in Institute
of Physics in Belgrade (Serbia). This transmitter was chosen because it provides the best quality
of the recorded signal owing to the high emission power of 800 kW and
suitable signal frequency for the location of the receiver, and a relatively short signal propagation path.
The latter property is important as it excludes significant variations in vertical stratification of parameters
in the ambient ionospheric plasma.
The amplitude and phase time variations of the recorded signal during
the low ionosphere responses to the considered perturbing flares are shown in the Fig.~\ref{Fig1::signali}
and their role in electron density calculations are explained in the next section.

As we can see from Fig.~\ref{Fig1::signali} the intensities of X-radiation decrease in time at the final part of the indicated time periods and they reach values existing prior to flare events. Within the same time period, the signal data show monotonous relaxations without visible additional changes. In the section that follows, it will be shown that zenith angles along the trajectory are small and do not change significantly during the observed time interval for the three considered cases. All these facts together allow us to apply the presented theoretical model to studied cases.

\section{Results}

As can be seen from Eq.~\ref{eq:13}, for the electron gain rate ${\cal G}(h,t)$ that tends to ${\cal G}_{Ly\alpha}(h)$ according to Eq.~(\ref{eq:2}), it is first necessary to determine the altitude and time dependencies of electron density $N(h,t)$.
In this work, in calculations we consider a relatively short observational period (around 30 s) and relatively short path of VLF signals characterized by small differences in solar zenith angle between locations of the transmitter (latitude 53.1 N and longitude 7.6 E) and the receiver (latitude 44.8 N and longitude 20.4 E) which are: 40.5   and 40.3,   54.0   and 51.4, and   66.4 and 62.4 for the Case I, Case II and Case III, respectively. The calculation given in \cite{han11}, \cite{mcr00} and \cite{fri01} for considered angles predict very small differences in parameters which yields small electron density changes. The relevant values are given in Table~\ref{table2} where data obtained by foregoing methods are considered separately. Here, we can see that along the trajectory the maximum deviations from the mean values relevant for the geographic locations of the transmitter and the receiver are very small, less then about 1\% for $\beta$, 2\% for $H'$ and 20\% for $N_e$.
This allows us to assume constant parameters in the unperturbed ionosphere which are then referent values for determined time dependencies of parameters during perturbations. Here, we present calculation that includes constant values $\beta$ = 0.3 km$^{-1}$ for sharpness and $H'$ = 74 km for the reflection height in unperturbed plasma, also used in previous works \citep {zig07,gru08,nin11,nin12a,nin12b}. These chosen values lie within presented relevant data for all considered angles.

  The time distributions of
  obtained parameters $H'(t)$ and $\beta(t)$} and electron density $N(h,t$) at different heights
  during the entire period of ionospheric response to the solar X flare are presented in Figs~\ref{Fig2::parametri} and~\ref{Fig3::N}.

  The obtained values of parameters $H'(t_{_{I_{max}}})$ and
  $\beta(t_{_{I_{max}}})$, and the electron densities $N(h,t_{_{I_{max}}})$
  at flare intensity peaks
   are in a good agreement with values presented in \cite{gru08}, \cite{tho05}
  and \cite{zig07}.

To calculate the electron density time derivative,
  we a suitable fitting function for the electron density time dependence in the relaxation period. We thus eliminate small (in our analysis)
  non important variations in $N(h,t)$ caused by other perturbers and consider only the general time
  tendencies. The fitted curves are shown in Fig.~\ref{Fig3::N} by solid lines,
  and the calculated time variations of the electron density time derivative are presented in the
  Fig.~\ref{Fig4::dNdt} where we can see that these time derivatives are negative
  and, after
  sufficiently long time,
  they approach zero at all considered altitudes. These two facts practically
  indicate the return of plasma to its initial state.

  Generally, the very slow variation of
  plasma characteristics
  at the end of considered time periods as seen,
   in Figs~\ref{Fig1::signali}-\ref{Fig4::dNdt}, enables the application of the presented theoretical model in
  the indicated time intervals
  and
  calculations of time dependencies of ${\cal G}(h,t)$ presented in Fig.~\ref{Fig5::G} for heights
  70, 75 and 80 km.
  Here, we can see that electron gain rates are practically time independent
  and we can assume that
  these values approximately determine the electron gain rates ${\cal G}_{Ly\alpha}(h)$ induced by the
  Ly-$\alpha$ line.


\begin{figure}
\begin{center}
\includegraphics*[width=8cm,angle=0]{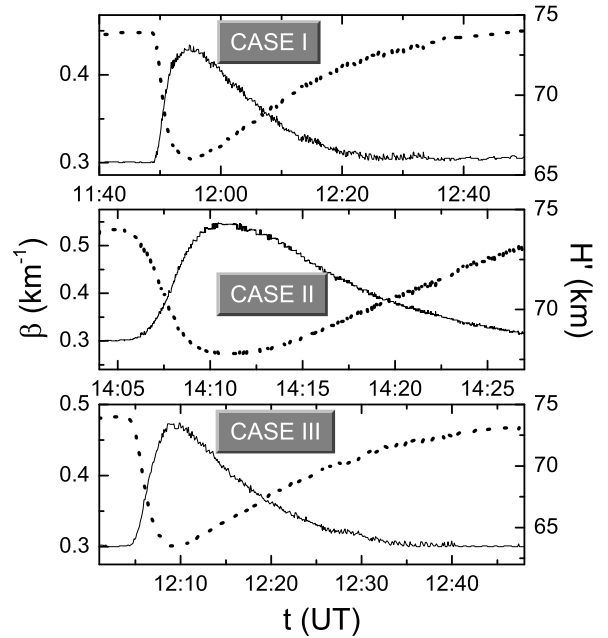}
\end{center}
\caption{Time dependencies of Wait's parameters during the considered flare activities. Solid and dotted lines are related to sharpness $\beta(t)$ and signal reflection height $H'(t)$, respectively.}
\label{Fig2::parametri}
\end{figure}

\begin{figure}
\begin{center}
\includegraphics*[width=8cm,angle=0]{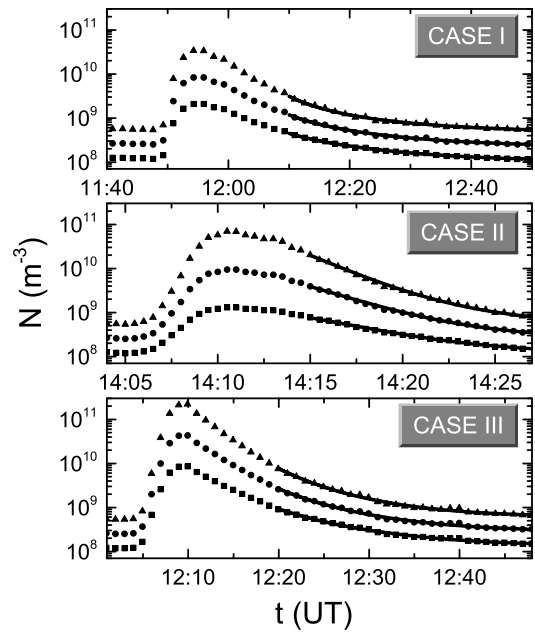}
\end{center}
\caption{Time dependencies of the electron density at altitudes 70 km (squares), 75 km (circles)
and 80 km (triangles) during the considered flare activities.}
\label{Fig3::N}
\end{figure}

\begin{figure}
\begin{center}
\includegraphics*[width=8cm,angle=0]{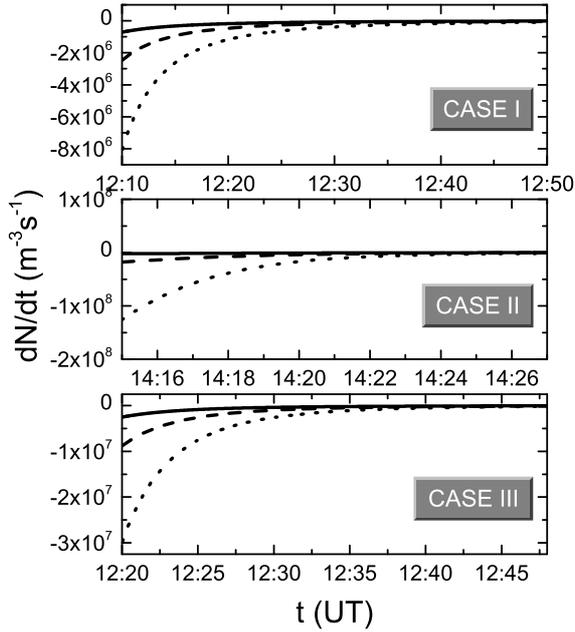}
\end{center}
\caption{Time dependencies of the electron density time derivative at
altitudes 70 km (solid lines), 75 km (dashed lines) and 80 km (dotted lines) during relaxation periods.}
\label{Fig4::dNdt}
\end{figure}

\begin{figure}
\begin{center}
\includegraphics*[width=8cm,angle=0]{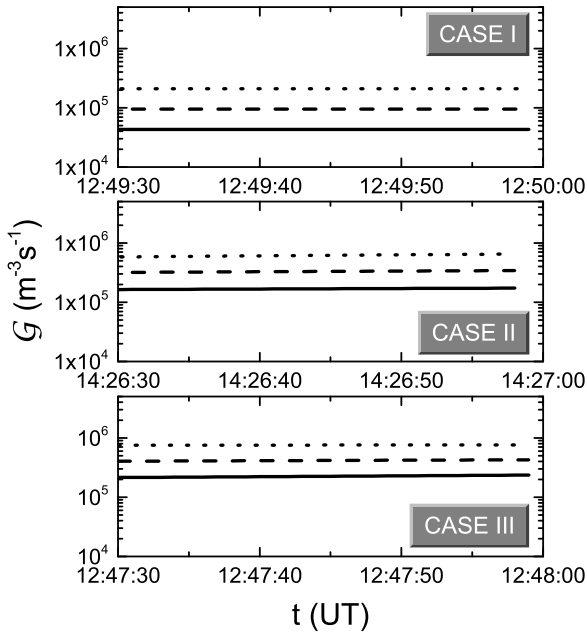}
\end{center}
\caption{Time dependencies of the electron gain rate ${\cal G}(h,t)$ at
altitudes 70 km (solid lines), 75 km (dashed lines) and 80 km (dotted lines) during the periods relevant to applying method.}
\label{Fig5::G}
\end{figure}

\begin{figure}
\begin{center}
\includegraphics*[width=9cm,angle=0]{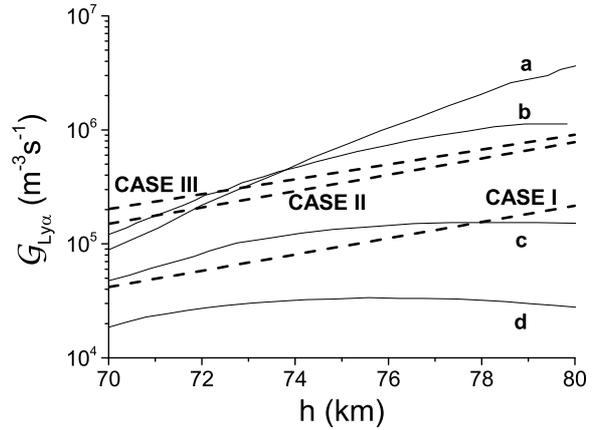}
\end{center}
\caption{Altitude dependencies of the electron gain rate
${\cal G}_{Ly\alpha}(h)$ for Case I, II and III, and their comparison with data
presented in \cite{mit77} (a), \cite{row72} (b), \cite{aik64} (c) and \cite{bou65} (d).}
\label{RADGLyalpha}
\end{figure}

The altitude distribution of ${\cal G}(h,t)$ can be found from Wait's equation for electron density Eq.~(\ref{eq:14}).
Namely, it yields the following ratio of electron densities $N(h_1,t)$ and $N(h_2,t)$ at heights $h_1$ and $h_2$,  respectively:

  \begin{equation}
    \label{eq:15}
    {
    \frac{N(t,h_1)}{N(t,h_2)} = e^{(\beta(t)-0.15)(h_1-h_2)}.
    }
  \end{equation}
If we further consider the ratio of the ionization rates at these altitudes, ${\cal G}(t,h_1)$ and ${\cal G}(t,h_2)$,
and taking into account that $\beta(t)$ is practically constant at the ends of considered time intervals,
Eqs~(\ref{eq:13}) and~(\ref{eq:15}) give:
  \begin{equation}
    \label{eq:16}
    {
    \frac{{\cal G}(t,h_1)}{{\cal G}(t,h_2)} \approx e^{(\beta(t)-0.15)(h_1-h_2)}.
    }
  \end{equation}
  The obtained values for ${\cal G}_{Ly\alpha}(75\rm{km})$ and $\beta(t)$ at the ends of the considered time intervals are inserted
  into the Eq.~(\ref{eq:16}) which yields the altitude distributions of the electron gain rate induced by the Ly-$\alpha$ line as shown in Fig.~\ref{RADGLyalpha}. Here, we can see that our model, when applied to the low ionospheric plasma perturbations induced by
  considered flares,
  gives the Ly-$\alpha$ photo-ionization rates that fall within the domain of values presented in \cite{mit77,row72,aik64} and \cite{bou65}. As we said in Introduction, the differences in the presented altitude variations are due to different plasma and propagation conditions, and due to different Ly-$\alpha$ intensities.

The error factor of our method, $EF$, defined as the ratio of ${\cal G'}_{Ly\alpha}(h)$ to ${\cal G}_{Ly\alpha}(h)$, where values of ${\cal G'}_{Ly\alpha}(h)$ are calculated from electron densities obtained by methods taking into account the Wait's parameters dependencies of the solar zenith angle, can be obtained from Eq.~(\ref{eq:13}) as:
    \begin{equation}
    \label{eq:16}
    {
    EF\equiv\frac{{\cal G'}_{Ly\alpha}(h)}{{\cal G}_{Ly\alpha}(h)} \approx \frac{N'(h)}{N(h)}.
    }
  \end{equation}
  We take here the electron density ratio obtained by two methods to be practically constant, i.e.  $N'(h,t)/N(h,t)\approx const$. This approximation is justified because of very small changes in plasma parameters within considered short time period.
  Table 2 shows comparisons of  $EF$ obtained by our method with those given in \cite{han11}, \cite{mcr00} and \cite{fri01}.
  Taking into account spreads of results shown in Fig.~\ref{RADGLyalpha} where the presented values differ by three orders of magnitude, we can conclude that our calculated error factor ($EF<4$ in all three considered cases) can be taken as relatively very small.
  From this fact and from comparisons with existing data shown in Fig.~\ref{RADGLyalpha}, we conclude that the presented calculating procedure of the photo-ionization rate induced by the Ly-$\alpha$ line radiation gives values which do not significantly depend on the method how the electron density is calculated, and which are consistent with values from the literature.

\section{Summary}

In this paper, we present a new theoretical procedure for calculations of the photo-ionization rate
in the terrestrial ionospheric D region induced by the Ly-$\alpha$ line radiation coming from the Sun.
This method is based on a slow time variation of plasma characteristics during the relaxation period
following a
large perturbation like that caused by a solar X flare. It requires observational data to determine
the electron density at the considered location,
which can be obtained
from the D region monitoring.

The theoretical results are applied to three cases with solar X flares occurred on May $5^{\rm th}$, 2010, February 18, 2011 and March 24, 2011,
where we used
the amplitude and phase time
variations of the VLF radio signal emitted by the DHO transmitter in Germany and recorded by
the AWESOME receiver located in Serbia
for low ionosphere monitoring.
The obtained
 altitude distributions of the photo-ionization rate
(order of magnitudes $10^4$ to $10^5$ m$^{-3}$s$^{-1}$) are in a good agreement with data existing in the literature.

\section*{Acknowledgments}
The present work was supported by the Ministry of
Education, Science and Technological Development of
the Republic of Serbia as a part of the projects no. III
44002, 176002 and 176004.

\bibliographystyle{elsarticle-harv}

\clearpage




%

\begin{landscape}
\begin{table}
\caption{Wait's parameters and electron densities at altitudes $h$ = 70, 75 and 80 km obtained by data from \cite{han11}, \cite{mcr00} and \cite{fri01} for solar zenith angles related to locations of the transmitter (A) and receiver (B) in considered cases. $\Delta_{\%}$ show deviations of these values from their mean values.}
\begin{tabular}{|c|c|c|c|c|c|c|c|}
\hline
  DATA	& {\multirow{2}{*}{CASE}} & & {\multirow{2}{*}{$H'$}} & {\multirow{2}{*}{$\beta$}} & \multicolumn{3}{ c| }{$N$} \\ \cline{6-8}
 SOURCE & & & & & 70 km & 75 km & 80 km \\
 \hline
 {\multirow{9}{*}{\cite{han11}}} &  {\multirow{3}{*}{I}} & A &	 73.2 km &	 0.385 km$^{\rm {-1}}$ &	$1.15\cdot 10^8$ m$^{\rm {-3}}$ &	 $3.72\cdot 10^8$ m$^{\rm {-3}}$ &	 $1.20\cdot 10^9$ m$^{\rm {-3}}$ \\
 \cline{3-8}
 &   & B	& 73.1 km & 0.385 km$^{\rm {-1}}$ & $1.19\cdot 10^8$ m$^{\rm {-3}}$ & $3.87\cdot 10^8$ m$^{\rm {-3}}$ & $1.25\cdot 10^9$ m$^{\rm {-3}}$ \\
\cline{3-8}
 &   &  \textbf{$\Delta_{\%}$} &  \textbf{0.07 \%} & 	 \textbf{0.00 \%} & 	 \textbf{1.92 \%}  & \textbf{1.92 \%} &	 \textbf{1.92 \%} \\
 \cline{2-8}
 &  {\multirow{3}{*}{II}} & A &	74.8 km &	0.381 km$^{\rm {-1}}$ &	$6.32\cdot 10^7$ m$^{\rm {-3}}$ &	$2.01\cdot 10^8$ m$^{\rm {-3}}$ &	$6.37\cdot 10^8$ m$^{\rm {-3}}$ \\ \cline{3-8}
 &   & B	& 75.2 km &	0.377 km$^{\rm {-1}}$ &	$5.54\cdot 10^7$ m$^{\rm {-3}}$ &	 $1.72\cdot 10^8$ m$^{\rm {-3}}$ &	$5.37\cdot 10^8$ m$^{\rm {-3}}$ \\
\cline{3-8}
 &   & \textbf{$\Delta_{\%}$} & \textbf{0.27 \%} &	\textbf{0.53 \%} &	 \textbf{6.57 \%} &	\textbf{7.57 \%} &	\textbf{8.56 \%} \\
\cline{2-8}

    &  {\multirow{3}{*}{III}} & A &	 73.9 km &	0.374 km$^{\rm {-1}}$ &	 $9.16\cdot 10^7$ m$^{\rm {-3}}$ &	$2.81\cdot 10^8$ m$^{\rm {-3}}$ &	$8.60\cdot 10^8$ m$^{\rm {-3}}$ \\
    \cline{3-8}
 &   & B	& 74.1 km & 	0.378 km$^{\rm {-1}}$ &	$8.36\cdot 10^7$ m$^{\rm {-3}}$ &	 $2.61\cdot 10^8$ m$^{\rm {-3}}$ &	$8.17\cdot 10^8$ m$^{\rm {-3}}$ \\
\cline{3-8}
 &   & \textbf{$\Delta_{\%}$} & \textbf{0.14 \%} &	\textbf{0.53 \%} &	 \textbf{4.56 \%} &	\textbf{3.56 \%} &	\textbf{2.56 \% }\\

\hline

  {\multirow{9}{*}{\cite{mcr00}}} &  {\multirow{3}{*}{I}} & A &	 71.6 km &	 $0.377 km^{\rm {-1}}$ &	$2.15\cdot 10^8$ m$^{\rm {-3}}$ &	 $6.70\cdot 10^8$ m$^{\rm {-3}}$ &	 $2.09\cdot 10^9$ m$^{\rm {-3}}$ \\
 \cline{3-8}
 &   & B	& 71.6 km &	0.376 km$^{\rm {-1}}$ &	$2.16\cdot 10^8$ m$^{\rm {-3}}$ &	 $6.68\cdot 10^8$ m$^{\rm {-3}}$ &	$2.07\cdot 10^9$ m$^{\rm {-3}}$ \\
\cline{3-8}
 &   &  \textbf{$\Delta_{\%}$} &  \textbf{0 \%} &	\textbf{0.13 \%} &	 \textbf{0.08 \%} &	\textbf{0.17 \%} &	\textbf{0.42 \%} \\
 \cline{2-8}

    &  {\multirow{3}{*}{II}} & A &	74.1 km &	0.324 km$^{\rm {-1}}$ &	 $1.04\cdot 10^8$ m$^{\rm {-3}}$ &	$2.49\cdot 10^8$ m$^{\rm {-3}}$ &	$5.94\cdot 10^8$ m$^{\rm {-3}}$ \\
    \cline{3-8}
 &   & B	& 74.8 km &	0.311 km$^{\rm {-1}}$ &	$8.85\cdot 10^7$ m$^{\rm {-3}}$ &	 $1.98\cdot 10^8$ m$^{\rm {-3}}$ &	$4.43\cdot 10^8$ m$^{\rm {-3}}$ \\
\cline{3-8}
 &   & \textbf{$\Delta_{\%}$} & \textbf{0.47 \%} &	\textbf{2.05 \%} &	 \textbf{8.20 \%} &	\textbf{11.42 \%} &	\textbf{14.61 \%} \\
\cline{2-8}

    &  {\multirow{3}{*}{III}} & A &	 72.7 km &	0.355 km$^{\rm {-1}}$ &	 $1.51\cdot 10^8$ m$^{\rm {-3}}$ &	$4.21\cdot 10^8$ m$^{\rm {-3}}$ &	$1.17\cdot 10^9$ m$^{\rm {-3}}$ \\
    \cline{3-8}
 &   & B	& 72.9 km &	0.351 km$^{\rm {-1}}$ &	$1.42\cdot 10^8$ m$^{\rm {-3}}$ &	 $3.89\cdot 10^8$ m$^{\rm {-3}}$ &	$1.06\cdot 10^9$ m$^{\rm {-3}}$ \\
\cline{3-8}
 &   & \textbf{$\Delta_{\%}$} & \textbf{0.14 \%} &	\textbf{0.57 \%} &	 \textbf{2.97 \%} &	\textbf{3.97 \%} &	\textbf{4.97 \%}\\

\hline

  {\multirow{9}{*}{\cite{fri01}}} &  {\multirow{3}{*}{I}} & A &	 71.0 km &	 0.263 km$^{\rm {-1}}$ &	$3.03\cdot 10^8$ m$^{\rm {-3}}$ &	 $5.33\cdot 10^8$ m$^{\rm {-3}}$ &	 $9.37\cdot 10^8$ m$^{\rm {-3}}$ \\
 \cline{3-8}
 &   & B	& 71.0 km &	0.263 km$^{\rm {-1}}$ &	$3.03\cdot 10^8$ m$^{\rm {-3}}$ &	 $5.33\cdot 10^8$ m$^{\rm {-3}}$ &	$9.37\cdot 10^8$ m$^{\rm {-3}}$ \\
\cline{3-8}
 &   &  \textbf{$\Delta_{\%}$} &  \textbf{0 \%} &	\textbf{0 \%} &	 \textbf{0 \%} &	\textbf{0 \%} &	\textbf{0 \%} \\
 \cline{2-8}

 &  {\multirow{3}{*}{II}} & A &	75.2 km &	0.226 km$^{\rm {-1}}$ &	$1.22\cdot 10^8$ m$^{\rm {-3}}$ &	$1.78\cdot 10^8$ m$^{\rm {-3}}$ &	$2.60\cdot 10^8$ m$^{\rm {-3}}$ \\
    \cline{3-8}
 &   & B	& 77 km &	0.217 km$^{\rm {-1}}$ &	$8.62\cdot 10^7$ m$^{\rm {-3}}$ &	 $1.21\cdot 10^8$ m$^{\rm {-3}}$ &	$1.68\cdot 10^8$ m$^{\rm {-3}}$ \\
\cline{3-8}
 &   & \textbf{$\Delta_{\%}$} & \textbf{1.18 \%} &	\textbf{2.03 \%} &	 \textbf{17.02 \%} &	\textbf{19.20 \%} &	\textbf{21.36 \%} \\
\cline{2-8}

    &  {\multirow{3}{*}{III}} & A &	 72.4 km &	0.245 km$^{\rm {-1}}$ &	 $2.19\cdot 10^8$ m$^{\rm {-3}}$ &	$3.52\cdot 10^8$ m$^{\rm {-3}}$ &	$5.66\cdot 10^8$ m$^{\rm {-3}}$ \\
    \cline{3-8}
 &   & B	& 72.9 km &	0.241 km$^{\rm {-1}}$ &	$1.96\cdot 10^8$ m$^{\rm {-3}}$ &	 $3.09\cdot 10^8$ m$^{\rm {-3}}$ &	$4.86\cdot 10^8$ m$^{\rm {-3}}$ \\
\cline{3-8}
 &   & \textbf{$\Delta_{\%}$} & \textbf{0.34 \%} &	\textbf{0.82 \%} &	 \textbf{5.54 \%} &	\textbf{6.54 \%} &	\textbf{7.53 \%}\\

\hline
 \end{tabular}
\label{table1}
\end{table}
\end{landscape}

\clearpage

\begin{table}
\caption{Error factors $EF$ for given method for implementation electron density values obtained by our procedure versus those calculated from data presented in \cite{han11}, \cite{mcr00} and \cite{fri01} for solar zenith angles related to the transmitter (A) and receiver (B) locations in considered cases and given in Eq. \ref{eq:16}.}
\begin{tabular}{|c|c|c|c|c|c|}
\hline
  DATA	& {\multirow{2}{*}{CASE}} & & \multicolumn{3}{ c| }{$EF$} \\ \cline{4-6}
 SOURCE & & & 70 km & 75 km & 80 km \\
 \hline
 {\multirow{6}{*}{\cite{han11}}} &  {\multirow{2}{*}{I}} & A &	1.01 &	 1.54 &	 2.35\\
 \cline{3-6}
 &   & B	&  0.97 &	1.48 &	2.27\\
 \cline{2-6}
 &  {\multirow{2}{*}{II}} & A &0.53 &	0.80 &	1.20	 \\
    \cline{3-6}
 &   & B	& 0.47 &	0.69 &	1.01\\
\cline{2-6}

    &  {\multirow{2}{*}{III}} & A &	0.77 &	1.12 &	1.62  \\
    \cline{3-6}
 &   & B	&  0.70 &	1.04 &	1.54\\

\hline

  {\multirow{6}{*}{\cite{mcr00}}} &  {\multirow{2}{*}{I}} & A &	1.82 &	 2.67 &	3.92 \\
 \cline{3-6}
 &   & B	&  1.82 &	2.66 &	3.89\\
 \cline{2-6}
 &  {\multirow{2}{*}{II}} & A &	0.88 &	1.00 &	1.12 \\
    \cline{3-6}
 &   & B	& 0.75 &	0.79 &	0.83 \\
 \cline{2-6}

    &  {\multirow{2}{*}{III}} & A &	1.27 &	1.68 &	2.21  \\
    \cline{3-6}
 &   & B	& 	1.20 &	1.55 &	2.00 \\

\hline

  {\multirow{6}{*}{\cite{fri01}}} &  {\multirow{2}{*}{I}} & A &	2.55 &	 2.12 &	 1.76 \\
 \cline{3-6}
 &   & B	& 2.55 &	2.12 &	1.76 \\
 \cline{2-6}
 &  {\multirow{2}{*}{II}} & A &1.03 &	0.71 &	0.49	\\
    \cline{3-6}
 &   & B	& 0.73 &	0.48 &	0.32 \\
\cline{2-6}

    &  {\multirow{2}{*}{III}} & A &	1.84 &	1.40 &	1.06  \\
    \cline{3-6}
 &   & B	& 1.65 &	1.23 &	0.91 \\

 \hline
 \end{tabular}
\label{table2}
\end{table}

\end{document}